\documentclass[10pt, journal, twoside]{IEEEtran}
\usepackage{cite} 
\usepackage{graphicx}
\usepackage{subfigure}
\usepackage{amssymb}
\usepackage{amsmath}
\usepackage{bm}
\usepackage{multirow}
\usepackage[OT1]{fontenc} 
\usepackage{algorithm}
\usepackage{algorithmic}
\usepackage[algo2e,ruled]{algorithm2e}
\usepackage{adjustbox}

\begin{document}
%
\title{Residual-Guided In-Loop Filter Using Convolution Neural Network}
%
%
%

\author{Wei Jia,
		Li Li,
        Zhu Li,
        Xiang Zhang,
        Shan Liu,
\thanks{W. Jia and Z. Li are with the Department of Computer Science and Electrical Engineering, University of Missouri-Kansas City, MO 64110, USA. Professor Zhu Li is the corresponding author (e-mail: wj3wr@umsystem.edu; zhu.li@ieee.org).

L. Li is with the Department of Electronic Engineering and Informance Science, University of Science and Technology of China. (email: lil1@ustc.edu.cn).

X. Zhang and S. Liu are with the Tencent Media Lab, 2747 Park Blvd, Palo Alto, CA 94306, USA. (email: xxiangzhang@tencent.com; shanl@tencent.com).
}
}
\markboth{} 
{JIA \MakeLowercase{\textit{et al.}}: Residual-Guided In-Loop Filter Using Convolution Neural Network}

\maketitle

\begin{abstract}
The block-based coding structure in the hybrid video coding framework inevitably introduces compression artifacts such as blocking, ringing, etc.
To compensate for those artifacts, extensive filtering techniques were proposed in the loop of video codecs, which are capable of boosting the subjective and objective qualities of reconstructed videos. 
Recently, neural network based filters were presented with the power of deep learning from a large magnitude of data.
Though the coding efficiency has been improved from traditional methods in High-Efficiency Video Coding (HEVC), the rich features and information generated by the compression pipeline has not been fully utilized in the design of neural networks.
Therefore, in this paper, we propose the  Residual-Reconstruction-based Convolutional Neural Network (RRNet) to further improve the coding efficiency to its full extent, where the compression features induced from bitstream in form of prediction residual is fed into the network as an additional input to the reconstructed frame.
In essence, the residual signal can provide valuable information about block partitions and can aid reconstruction of edge and texture regions in a picture.
Thus, more adaptive parameters can be trained to handle different texture characteristics.
The experimental results show that our proposed RRNet approach presents significant BD-rate savings compared to HEVC and the state-of-the-art CNN-based schemes, indicating that residual signal plays a significant role in enhancing video frame reconstruction. 

\end{abstract}

\IEEEpeerreviewmaketitle

\section{Introduction}
\label{Sec::introduction}
Advanced Video Coding (H.264/AVC) \cite{wiegand2003overview}, High-Efficiency Video Coding (H.265/HEVC) \cite{sullivan2012overview} are existing popular video coding standards. Versatile Video Coding (VVC) \cite{Bross2019} is the emerging next-generation standard under the development of the Moving Pictures Expert Group (MPEG).
These video coding standards adopt the so-called hybrid coding frameworks, where the major procedures include prediction, transform, quantization, and entropy coding.
In the hybrid coding framework, a video frame is partitioned into non-overlapping coding blocks.
These blocks form the basis coding units (CU), prediction units (PU), transform units (TU), etc.
A block-based coding scheme is hardware-friendly and easy to implement.
It also lends itself to useful coding functionalities such as parallelization. 

However, block-wise operation inevitably introduces video quality degradation near the block boundaries, known as block artifacts.
Beyond that, coarse quantization is another major factor in causing video quality degradation, especially at the regions with sharp edges known as the ringing artifacts.
This ripple phenomenon induces poor visual quality and leads to a bad user-experience \cite{lim2011ringing}.
Given this, extensive in-loop filters have been proposed to compensate for artifacts and distortions in video coding. 
The in-loop filters can be classified into two categories based on whether the deep learning techniques are used.

The first category is traditional signal processing based methods, including Deblocking Filter (DF) \cite{liu2007post, norkin2012hevc}, Sample Adaptive Offset (SAO) \cite{1Fu2010, 2Fu2011, 3Fun2011}, Adaptive Loop Filter (ALF) \cite{tsai2013adaptive}, non-local in-loop filter \cite{ma2016nonlocal} and many others.
DF can reduce blocking artifacts at PU and TU boundaries. 
SAO compensates the pixel-wise residuals by explicitly signaling offsets for pixel groups with similar characteristics. 
ALF is essentially a Wiener filter where the current pixel is filtered as a linear combination of neighboring pixels.
The three filters mentioned above are based on neighbor-pixel statistics. In contrast, the non-local in-loop filter takes advantage of non-local similarities in natural images.

Traditional methods improve the video quality with relatively low complexity.
Therefore, they have been successfully applied in video coding standards. 
Recently, however, the deep learning based in-loop filters have been proposed to achieve further improvements \cite{zhang2018residual, lu2018deep, jia2019content}. 
One type of CNN utilizes the principle of the Kalman filter to construct a deep learning filter.
Another type of CNN consists of the highway or content-aware block units to achieve flexibility. 
People have realized that these deep learning based schemes have at least two benefits from traditional methods. 
One is that non-linear filtering operations are involved in the system. 
It is critical to capture and compensate for the distortions caused by codecs because these coding distortions are essentially non-linear by themselves. 
Another benefit is that deep learning can learn features from a large amount of data automatically, which would be more efficient than handcraft features.
Though the coding efficiency has been improved from traditional methods in HEVC, the coding information has not been fully utilized in the design of neural networks.
In \cite{he2018enhancing}, the authors proposed to utilize partition information in the design of neural networks, indicating introducing more coding information can benefit the overall performance.

Motivated by these, we propose a novel in-loop algorithm by introducing the residual signal to the network and devising two sub-networks for residual and reconstruction signals, respectively. 
They are the Residual Network and the Reconstruction Network. 
The major contributions of this work are three-fold:
\begin{itemize}
\item First, we supply the residual signal as the supplementary information and feed it into the neural network in pair with the reconstructed frame. 
To the best of our knowledge, this is the first work that utilizes the residual signal to devise an in-loop filter for video coding.
\item Second, the network structure is carefully designed for the dual-input CNN to utilize the underlying features in different input channels fully.
The residual blocks are used for Residual-Network. 
A hierarchical autoencoder network with skip connections is used for Reconstruction-Network.
\item Third, extensive experiments have been conducted to compare with existing algorithms to demonstrate its effectiveness of the proposed scheme. 
Throughout analyses are provided to give more insights into the problem based on the experimental results.
\end{itemize}

Note that a residual introduced deblocking method has been proposed in our previous work \cite{jia2020Residual}.
This paper provides more motivation, analysis, experimental results, and comparison of related works on the residual-based loop filter.
Additionally, in order to validate the efficiency of our RRNet design, we recurs more three inputs-based methods for comparison.
The experimental results show that the customized Residual Network and Reconstruction Network is significantly beneficial for bitrate savings. 

We organize the remainder of this paper as follows. 
In Section~\ref{Sec::related work}, we describe related works. 
Section~\ref{Sec::proposed algorithm} introduces the proposed RRNet approach.
In Section~\ref{Sec::experimental results}, we report and analyze the experimental results.
Finally, Section~\ref{Sec::conclusion} summarizes this paper and discusses future works.

\section{Related Work}
\label{Sec::related work}

In this section, we briefly review the prior works related to loop filters of video coding, including the traditional signal processing based methods and deep learning based methods.
\subsection{Traditional signal processing based methods}
Relying on the signal processing theory, the following in-loop filter methods have been proposed.

\begin{itemize}
\item [1)] Deblocking Filter (DF).
List \emph{et al.} \cite{list2003adaptive} devised the first version of an adaptive deblocking filter, which was adopted by H.264/AVC standard.
It depressed distortions at block boundaries by applying an appropriate filter. 
Zhang \emph{et al.} \cite{6166366} proposed a three-step framework considering task-level segmentation and data-level parallelization to efficiently parallelize the deblocking filter.
Tsu-Ming \emph{et al.} \cite{liu2007post} then proposed a high-throughput deblocking filter.
In HEVC, Norkin \emph{et al.} \cite{norkin2012hevc} designed a DF with lower complexity and better parallel-processing capability.
Li \emph{et al.} \cite{8085172} provided deblocking with a shape-adaptive low-rank before preserving edges well and an extra before restoring the lost high-frequency components.
\item [2)] Sample Adaptive Offset (SAO) \cite{fu2012sample}.
Chien and Karczewicz proposed an adaptive loop filtering technique \cite{Chien2009} based on the Laplacian energy and classifications of the reconstructed pixel value.
This approach obtains obvious performance improvements but with high complexity.
Ken \emph{et al.} \cite{Ken2010} designed an extrema correcting filter (EXC) and a boundary correcting filter (BDC).
Huang \emph{et al.} \cite{Huang2010} developed a picture-based boundary offset (PBO), picture-based border offset (PEO) and picture-based adaptive constraint (PAC).
Fu \emph{et al.} \cite{1Fu2010,2Fu2011} devised an algorithm that can adaptively select the optimal pixel-classification method.
However, computational complexity is still very high.
To address this, Fu and Chen \emph{et al.} \cite{3Fun2011} proposed a sample adaptive offset (SAO) method, which was finally adopted by HEVC.
It provides a better trade-off between performance and complexity.
\item [3)] Adaptive Loop Filter (ALF).
Tsai \emph{et al.} \cite{tsai2013adaptive} proposed the ALF method to decrease the mean square error between original frames and decoded frames by Wiener-based adaptive filter.
The filter coefficients are trained for different pixel regions at the encoder.
The coefficients are then explicitly signaled to the decoder.
Besides, ALF activates the filter at different regions by signaling control flags.
\item [4)] Non-local Mean Models.
The non-local mean methods improve the efficiency of in-loop filters as well.
To suppress the quantization noise optimally and improve the quality of the reconstructed frame, Han \emph{et al.} \cite{han2014quadtree} proposed a quadtree-based non-local Kuan’s (QNLK) filter. 
Ma \emph{et al.} \cite{ma2016nonlocal} proposed the group-based sparse representation with image local and non-local self-similarities. 
This model lays a solid groundwork for the in-loop filter design.
Zhang \emph{et al.} \cite{zhang2016low} utilized image non-local prior knowledge to develop a loop filter by imposing the low-rank constraint on similar image patches for compression noise reduction.
\end{itemize}

\begin{figure}[tbp]
\centering
\begin{minipage}[b]{.49\linewidth}
  \centering
  \includegraphics[width=\linewidth]{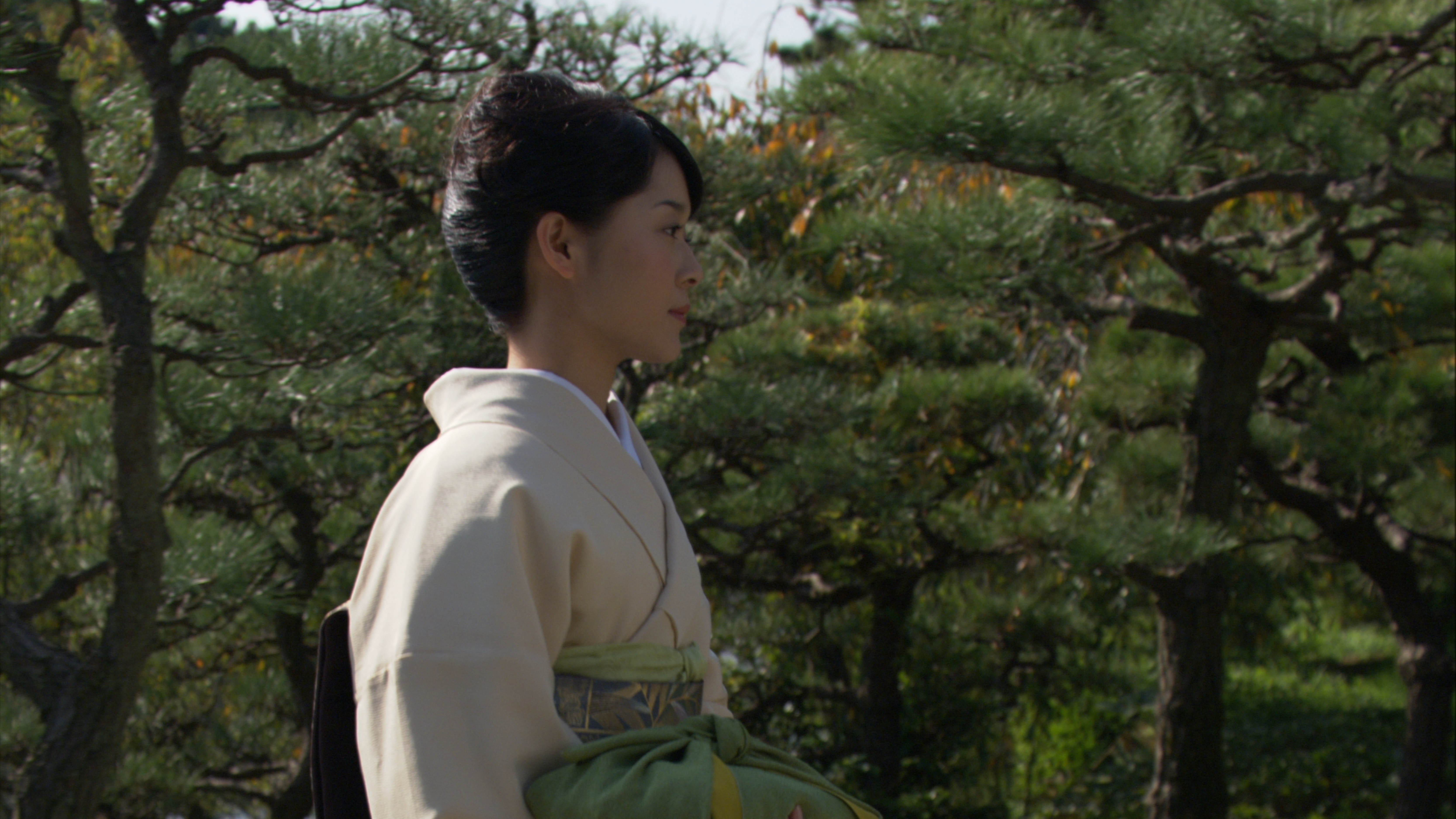}
  \centerline{(a) Ground Truth}\medskip
\end{minipage}
\begin{minipage}[b]{.49\linewidth}
  \centering
  \includegraphics[width=\textwidth]{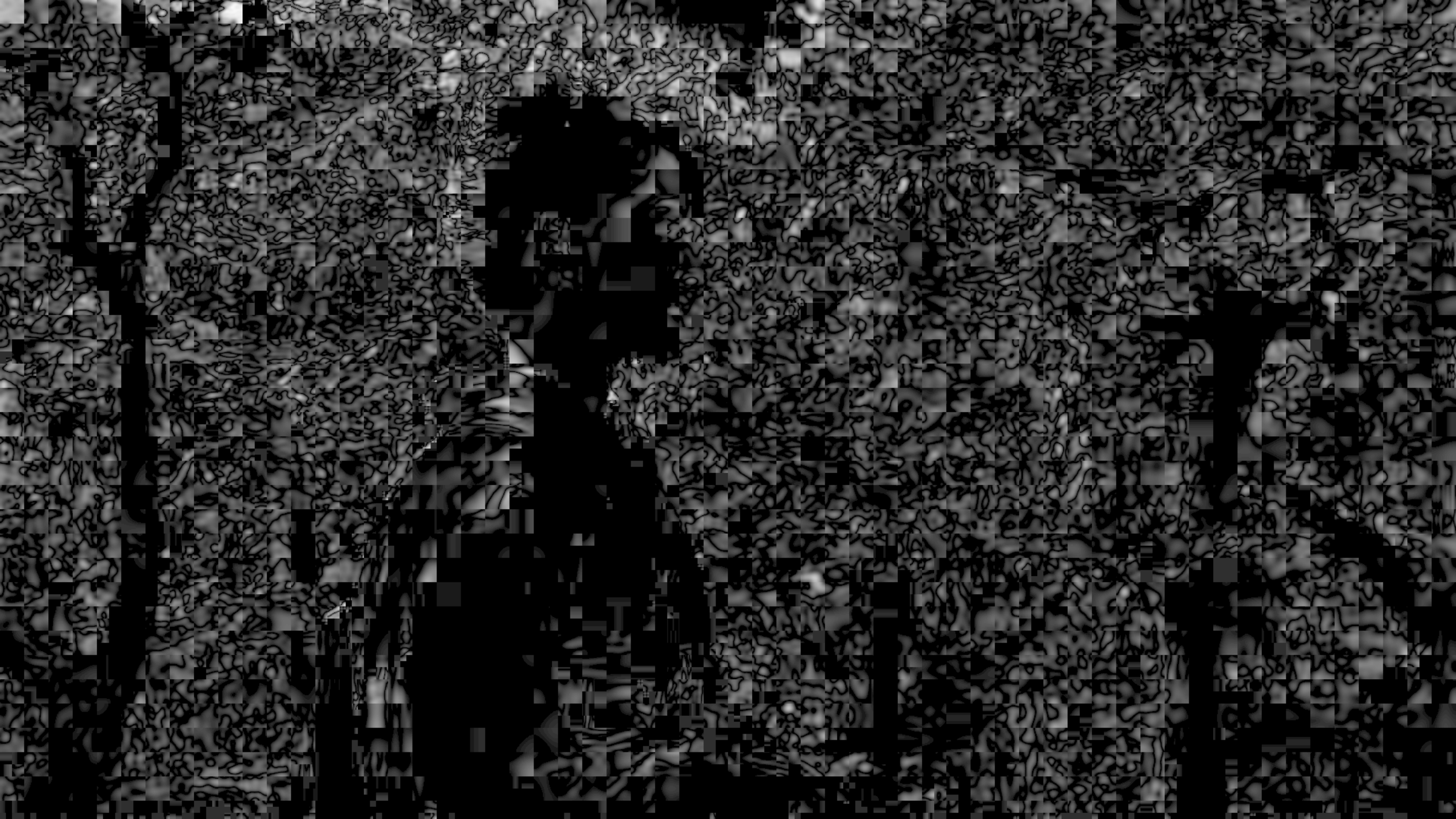}
  \centerline{(b) Residual}\medskip
\end{minipage}
\caption{Typical example of the Kimono residual under $QP37$ with intra mode. 
The color has been adjusted for clear viewing.
The inverse transformed residual signal provides the comprehensive partition information of the transforming units.
It is obvious to see the $32\times32$, $16\times16$, $8\times8$, and $4\times4$ partition blocks of TU in the residual.
For instance, the shapes of the woman's body and tree trunks are more easily discernable.
Meanwhile, the residual contains a large amount of dense, detailed textures. 
For example, we can see many needle leaves on the trees. This information can help to augment the considerable variation in some areas of the reconstruction.
}
\label{Fig::Example}
\end{figure}

\begin{figure*}[tbp]
\begin{center}
\centering
\includegraphics[width=1.0\textwidth]{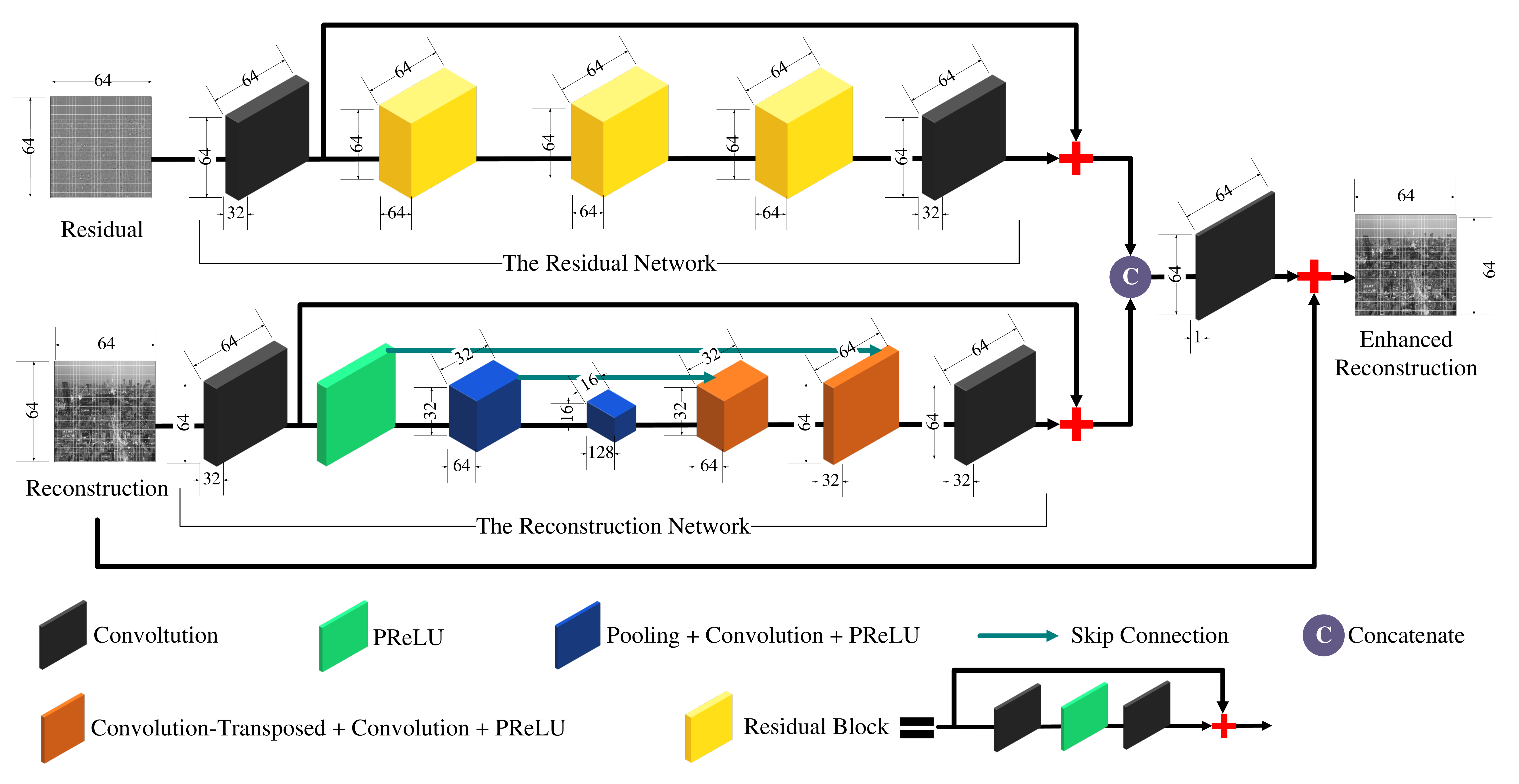}
\caption{RRNet with sub-networks of both the Residual Network and the Reconstruction Network.
Residuals are fed into the Residual Network to provide the TU partition information and the detailed textures information. 
The Residual Network relies on residual blocks to learn features effectively with residual learning.
We feed the reconstruction into the Reconstruction Network.
The Reconstruction Network executes the downsampling and upsampling strategy to patch up the local and global information. 
This enhances reconstruction quality and aids with the residual learning approach.
}
\label{Fig::HFAF}
\end{center}
\end{figure*}

\subsection{Deep learning based methods}
\label{subsec::cnnMethods}
Recently, the deep learning based in-loop filters have been proposed.
For images, Dong \emph{et al.} \cite{dong2015compression} designed a compact and efficient model, known as Artifacts Reduction Convolutional Neural Networks (AR-CNN).
This model was effective for reducing various types of coding artifacts.
Kang \emph{et al.} \cite{7109159} propose to learn sparse image representations for modeling the relationship between low-resolution and high-resolution image patches in terms of the learned dictionaries for image patches with and without blocking artifacts, respectively.
Wang \emph{et al.} \cite{wang2016d3} devised a Deep Dual-Domain ($D^3$) based fast restoration framework to recover high-quality images from JPEG compressed images.
The $D^3$ model increased the large learning capacity of deep networks.

For videos, Xue \emph{et al.} \cite{xue2017video} proposed the task-oriented flow (TOFlow), where a motion representation was learned for video enhancement.
Tao \emph{et al.} \cite{tao2017detail} proposed a sub-pixel motion compensation (SPMC) model, which has shown its efficiency in video super-resolution applications.
In the framework of video coding, Dai \emph{et al.} \cite{dai2017convolutional} designed a Variable-filter-size Residual-learning CNN (VRCNN) that achieved $4.6\%$ bit-rate gain.
Yang \emph{et al.} \cite{yang2017decoder,yang2018enhancing} developed the Quality Enhancement Convolutional Neural Network (QE-CNN) method in HEVC.
With the residual learning \cite{he2016deep}, Wang \emph{et al.} \cite{wang2018dense} designed the dense residual convolutional neural network (DRN), which exploits the multi-level features to recover a high-quality frame from a degraded one.
Other CNN-based video compression works, including \cite{jia2017spatial, park2016cnn, wang2017novel} pushed the horizon of in-loop filtering techniques as well. 
Most recently, 
Zhang \emph{et al.} \cite{zhang2018residual} devised the residual highway convolutional neural network (RHCNN) in HEVC.
Lu \emph{et al.} \cite{lu2018deep} modeled loop filtering for video compression as a Kalman filtering process.
Jia \emph{et al.} \cite{jia2019content} proposed a content-aware CNN based in-loop filtering for HEVC.
However, most of these frameworks are designed for one specific restoration task.
To address this issue, Jin \emph{et al.} \cite{8820082} proposed a flexible deep CNN framework that exploits the frequency characteristics of different types of artifacts.


\begin{figure}[tbp]
\centering
\begin{minipage}[b]{.49\linewidth}
  \centering
  \includegraphics[width=\linewidth]{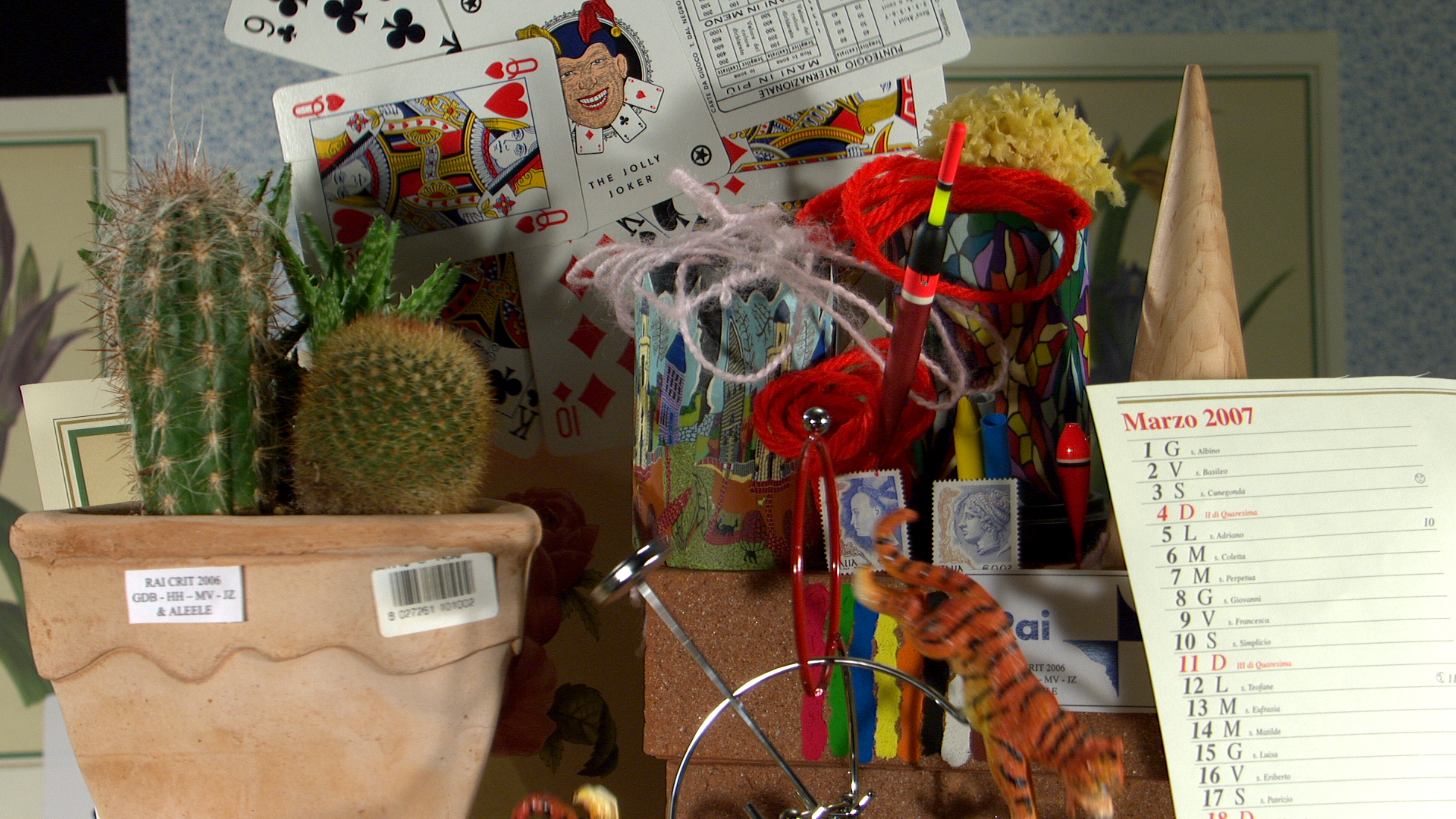}
  \centerline{(a) Cactus Ground Truth}\medskip
\end{minipage}
\begin{minipage}[b]{.49\linewidth}
  \centering
  \includegraphics[width=\linewidth]{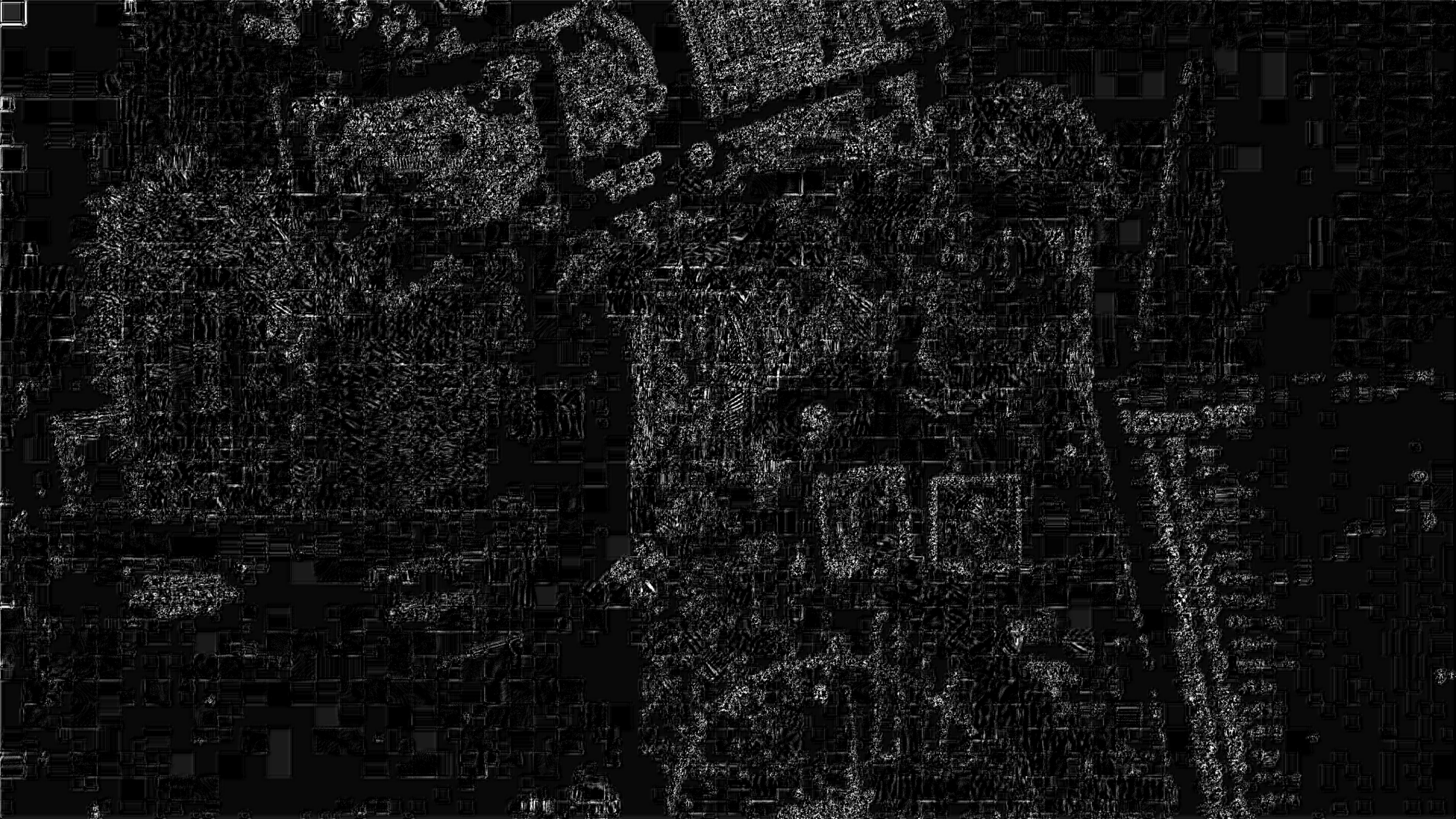}
  \centerline{(b) Cactus Residual Feature Map}\medskip
\end{minipage}
\begin{minipage}[b]{.49\linewidth}
  \centering
  \includegraphics[width=\linewidth]{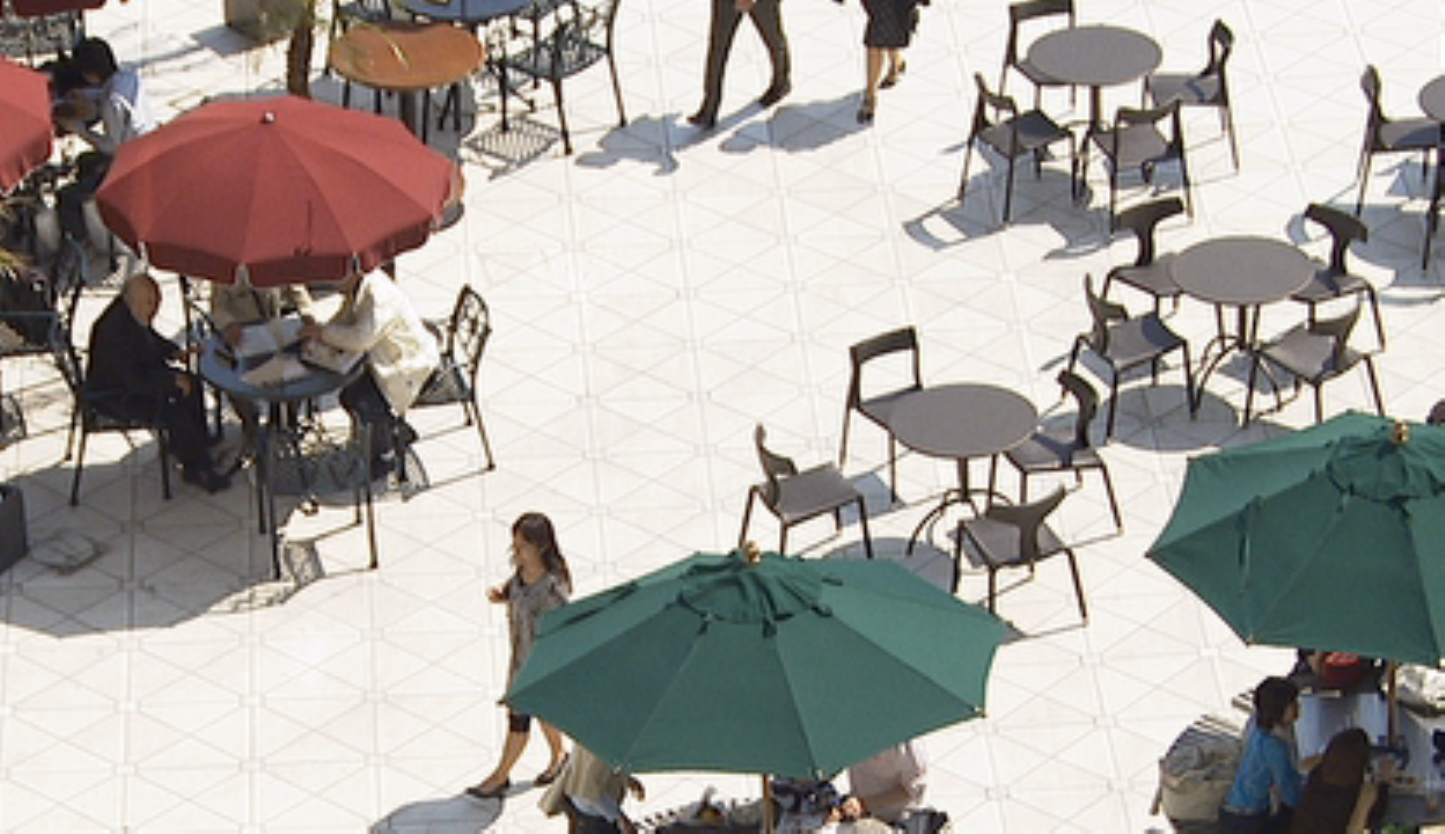}
  \centerline{(c) BQSquare Ground Truth}\medskip
\end{minipage}
\begin{minipage}[b]{.49\linewidth}
  \centering
  \includegraphics[width=\linewidth]{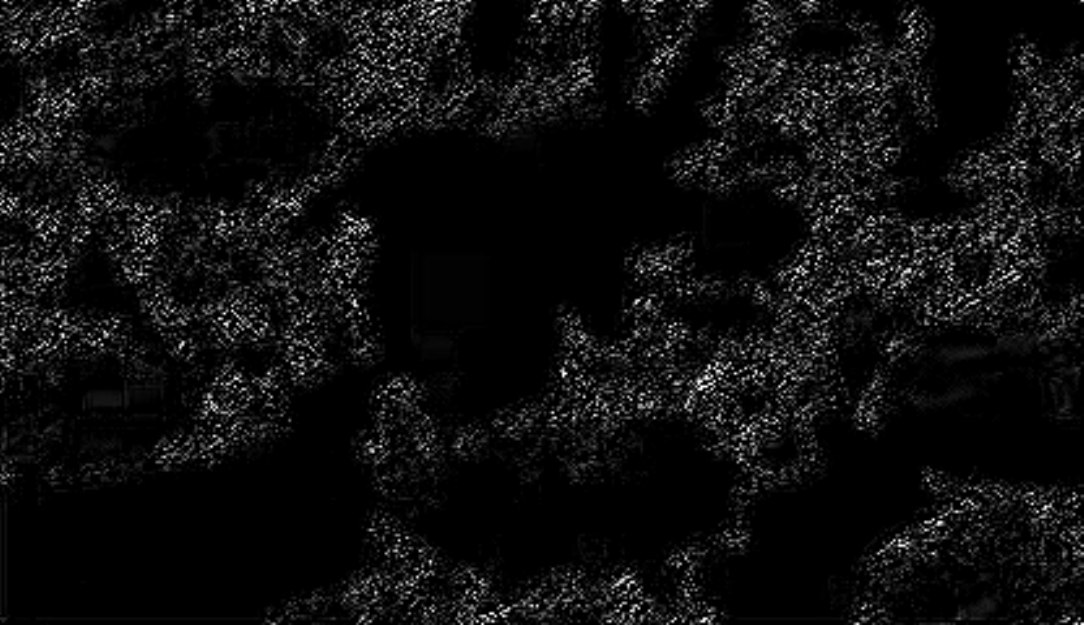}
  \centerline{(d) BQSquare }
  \centerline{ Residual Feature Map }
\end{minipage}
\caption{Residual feature maps of Cactus and BQSquare derived from the Residual Network of RRNet under $QP37$.
The residual features of Cactus with abundant context including pokers, calender and metal circle demonstrates its prominent contribution for enhancing the quality of the video frame.
The residual features of BQSquare which are a flat example show a great amount of details involving chairs and tables.
}
\label{Fig::featureMap}
\end{figure}

\begin{figure*}[tbp]
\begin{center}
\centering
\includegraphics[width=1.0\textwidth]{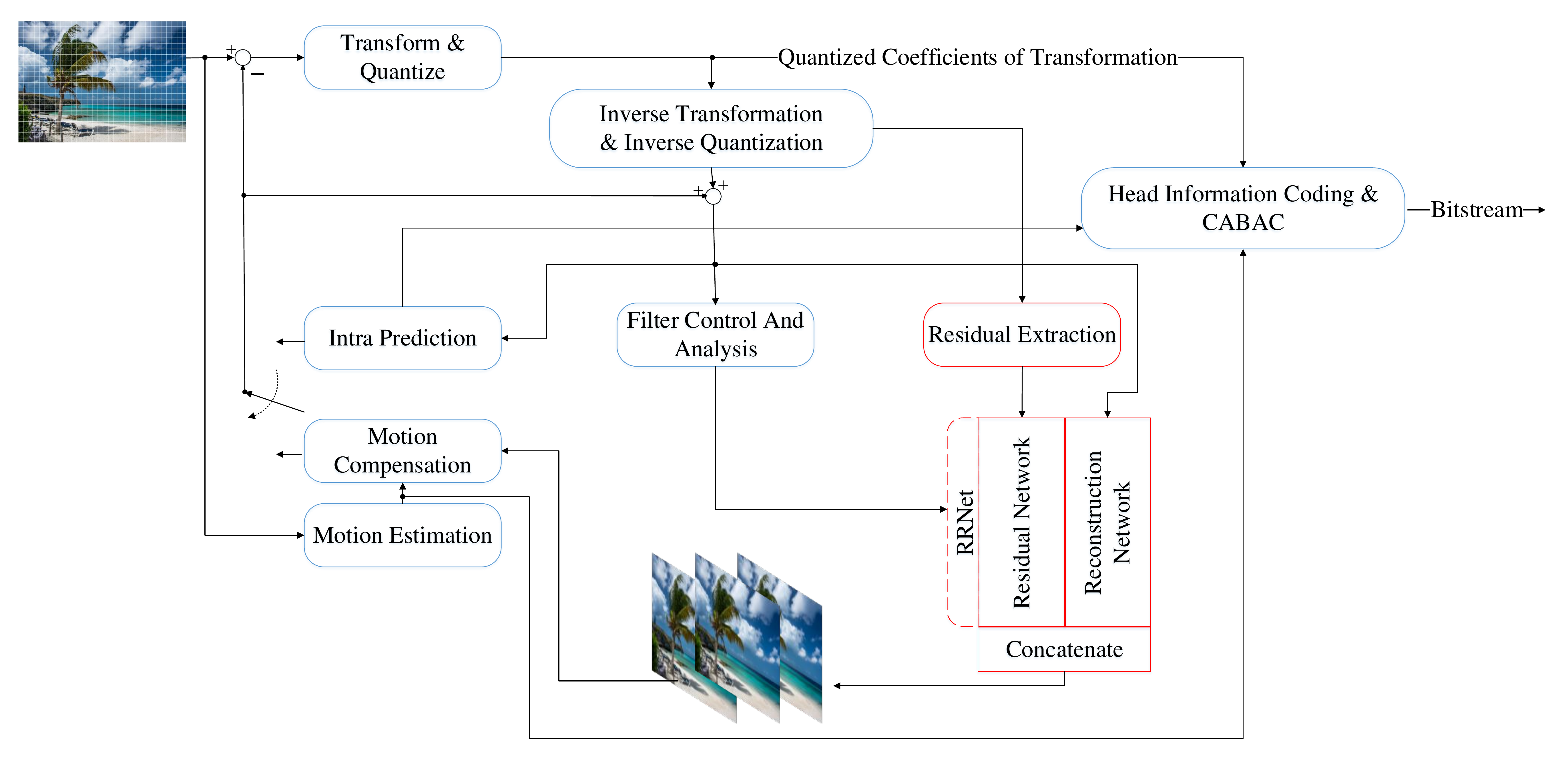}
\caption{The location of RRNet embedded in HEVC.
We insert the RRNet into HEVC as an in-loop method. The RRNet would input residual from extracting module and reconstruction into the Residual Network and the Reconstruction Network, respectively. The RRNet is executed instead of DF and SAO filters.
}
\label{Fig::codingFrame}
\end{center}
\end{figure*}

The aforementioned deep learning methods only took the reconstructed low-quality video frame as input.
However, the coding information was not efficiently utilized.
To better use coding information, Lin and He \emph{et al.} \cite{lin2019partition, he2018enhancing} proposed a partition-masked CNN, where the block partition information was utilized for improving the quality of the reconstructed frames.
It has shown additional improvements in terms of coding efficiency over the reconstruction-only methods.



\section{Proposed algorithm}
\label{Sec::proposed algorithm}

This section will discuss the proposed RRNet scheme in detail, including a more in-depth discussion on the architecture of the RRNet, loss function, dataset, and training process.

\begin{table}[tbp]
\caption{The Residual Network Parameters of conv layers}
\label{tab::resi}
\center
\begin{tabular}{l|c|c|c|c}
\hline
Layers                     &   Kernel Size       &   Feature maps   &   Stride & Padding\\
                           & & Number & & \\
\hline
\multirow{1}{*}{Conv 1}         &   $3 \times 3 $     &   $32$  &      $1$ & $1$         \\
\hline
Residual Block 1   &   $3 \times 3 $ &   $64$ & $1$ & $1$ \\ 
          (2 convs) & & & & \\
\hline 
Residual Block 2  &   $3 \times 3 $      &   $64$ & $1$ & $1$ \\
          (2 convs) & & & &\\
\hline
Residual Block 3  &   $3 \times 3 $      &   $64$ & $1$ & $1$  \\ 
 (2 convs) & & & & \\
\hline
\multirow{1}{*}{Conv 8}         &   $3 \times 3 $     &   $32$  &      $1$ & $1$                \\
\hline
\end{tabular}
\end{table}

\begin{table*}[tbp]
\caption{The Reconstruction Network Parameters of Conv And Transposed Conv Layers}
\label{tab::AFParas}
\center
\begin{adjustbox}{max width=\textwidth}
\begin{tabular}{c|c|c|c|c|c|c|c|c}
\hline
\multirow{1}{*}{Type of Layer}         & Conv1 & Conv2 & Conv3 & Transposed Conv1 & Conv4 & Transposed Conv2 & Conv5 & Conv6           \\
\hline
\multirow{1}{*}{Kernel Size}         & $3 \times 3$ & $3 \times 3$ & $3 \times 3$ & $2 \times 2$ & $3 \times 3$ & $2 \times 2$ & $3 \times 3$ & $3 \times 3$           \\
\hline
\multirow{1}{*}{Feature Map Number}         & $32$ & $64$ & $128$ & $64$ & $64$ & $32$ & $32$ & $32$      \\
\hline
\multirow{1}{*}{Stride}         & $1$ & $1$& $1$& $2$& $1$& $2$& $1$& $1$               \\
\hline
\multirow{1}{*}{Padding}         &   $1$ & $1$& $1$& $0$& $1$& $0$& $1$& $1$                \\
\hline
\end{tabular}
\end{adjustbox}
\end{table*}

\subsection{Architecture of the proposed RRNet framework}
\label{Subsec::archiRRNet}

Fig.~\ref{Fig::HFAF} shows the overall architecture of the proposed RRNet framework.
The proposed RRNet framework includes two sub-networks: the reconstruction network and the residual network.
The reconstruction network uses the reconstruction as input and derives reconstruction feature maps from the input.
The residual network uses the residual as input and derives residue feature maps from the input.
The feature maps derived from the two sub-networks are concatenated together and used as the input of the last convolutional layer.
In addition, we use the residual learning method that learns the difference between the input and the label to accelerate the training process.

As explained in the last paragraph, both the reconstruction and residual are utilized as the inputs of the proposed network.
Applying the reconstruction as input is the same as most existing works since our target is to enhance the reconstruction.
However, why the residual is used as the other input for our proposed RRNet network?

First, we believe that the residual can provide accurate transform unit (TU) partitions and great textures beneficial for the enhancement. 
Fig.~\ref{Fig::Example} gives a typical example of the residual from the sequence Kimono. 
We can see clear TU boundaries from the residual figure.
As we know, the basic unit of encoding the residual is a TU. 
Each TU transforms and quantizes independently. 
Therefore, it is more probable to have severe artifacts in the block boundary than the block center. 
The TU boundary information is a good indicator that implicates where the distortion is more severe and guides the network to learn more distinct features. 
In addition, we can see from the residual frame in Fig.~\ref{Fig::Example} that, within each TU, the texture information is still visible.
They can illustrate the body shapes of the girl and tree trunks clearly. 
This texture information also contributes to the reconstruction enhancement.

Second, the residual signal suffers from frame prediction accuracy, most notably in the areas where the residual contains non-zero values.
This essentially means that the encoder does not accurately predict the regions where the residual values are large.
Accordingly, the residual is beneficial for the CNN learning process, especially in areas where the residual contains non-zero values.
From the extracted residual feature maps as shown in Fig.~\ref{Fig::featureMap}, we can see that the residual signal is useful for improving the capability of the CNN to learn sharp edges and complex shape information that would otherwise be missed by the encoder.

In addition to introducing the residual as the dual input, we can also see from Fig.~\ref{Fig::HFAF} that we use different sub-networks for the reconstruction and residual.
As we know, the characteristics of the reconstruction and residual are different.
The residual is more sensitive, while the reconstruction consisting of residual and prediction contains more global information.
We should design specific sub-networks to optimize the features derived from various inputs and improve the reconstructed frame quality.
A detailed introduction of the two sub-networks will be described in detail in the next two subsections.

To give a better illustration of how we embed the above-introduced framework in HEVC, we give a modified HEVC encoder in Fig.~\ref{Fig::codingFrame}.
We replace the deblocking and SAO filters using the proposed RRNet framework.
The output frame from our framework will be used as a reference for the to-be-encoded frames in the future.
Note that in the proposed RRNet framework, we need to extract the residual from the bitstream in addition to the reconstruction.

\subsection{Design of the Residual Network}
\label{Subsec::hfcnn}

We develop a Residual Network consisting of several residual blocks \cite{he2016deep} to adapt to the residual features.
The residual block could effectively keep the residual features and the gradient information on the shallow layers.
Therefore, the proposed Residual Network can derive the distinct features from the residual frame.
Considering the complexity, we use only $8$ convolutional layers to derive the residual features.
Because the network consisting of residual blocks \cite{he2016deep} could effectively keep the residual features and the gradient information on the shallow layers \cite{veit2016residual}, we adopt the residual block as the basic unit of our Residual Network.

The network based on the residual blocks brings apparent advantages.
In the residual blocks based network, the collection of multiple routes substitutes the simple sole route.
Based on the multiple routes property, because of the independence of the routes in the residual block-based network, this uncorrelated property enhances the canonical effect of the Residual Network. 
Because the contributions for the gradient information are mainly from the shallow layers, adding the weights of the short routes could effectively prevent from vanishing gradient.

In Fig.~\ref{Fig::HFAF}, the upper pathway shows the detailed architecture of our proposed Residual Network.
Table~\ref{tab::resi} shows the convolutional layers configurations.
The Residual Network includes three residual blocks consisting of six convolutional layers and two convolutional layers at the beginning and end.
We set the Kernel Size for each convolutional layer as $3 \times 3$, the Feature Map Number as $32$, Stride as $1$, and Padding as $1$.
As the Parametric Rectified Linear Unit (PReLU) \cite{he2015delving} has been demonstrated to be more effective than the ReLU, we employ it as the activation function in the Residual Network.
We compute the feature maps of the Residual Network as follows:


\begin{equation}
\left\{
            \begin{array}{ll}
             F^{res}_i(x) = A(W_i * F^{res}_{i-1}(x) + B_i), i\in\{2,4,6,8\} & \\
                        F^{res}_j(x) = A(W_j * F^{res}_{j-1}(x) + B_j) + \ F^{res}_{j-2}(x), \  j\in\{3,5,7\} &  
            \end{array}
\right.
\label{Eq::resi}
\end{equation}
where $x$ denotes the input of residual, $A$ is the activation function, $W_i$ and $B_i$ are the weights and bias matrices respectively.

\subsection{Design of the Reconstruction Network}
\label{Subsec::afcnn}

Simultaneously, we consider the reconstruction signal as the other input.
Therefore, we design a Reconstruction Network containing several downsampling and upsampling pairs to learn the reconstruction features.
The Reconstruction Network adopts the classic autoencoder architecture \cite{chen2016variational, hinton2006reducing} with the skip connection concatenating the encoder and decoder parts \cite{ronneberger2015u}.
In this way, the reconstruction network can recover the global information and details as much as possible.

The Reconstruction Network has the following advantages.
On the encoder side, downsampling the reconstruction helps extract more useful reconstruction features of low space dimensions. 
Based on the downsampling operation, upsampling the small reconstruction features helps derive the more extensive reconstruction features on the decoder side. 
The skip connection concatenating the reconstruction features from the encoder side could help the decoder to recover the global and detailed information of the reconstruction.

In Fig~\ref{Fig::HFAF}, the lower pathway shows the detailed structure of our proposed Reconstruction Network.
We adopt the pooling and transposed convolutional layer to perform downsampling and upsampling, respectively.
In the encoder phase, downsampling reduces the redundancy effectively in the reconstruction and keeps useful information.
However, it may cut the global context as well.
Hence, we execute the upsampling in the decoder phase to propagate the global information of the reconstruction to the next convolutional layer.
Next, in the skip connection phase, we concatenate the concentrated reconstruction features from the encoder to the upsampling reconstruction features from the decoder.
This is to provide the network with both the brief features and global context in the reconstruction. 
The Reconstruction Network is a difference learning network as well.
Table~\ref{tab::AFParas} shows the detailed configurations.
For the convolutional layers, we set the Kernel Size to $3 \times 3$, Stride to $1$, Padding to $1$, Feature Map Number to $32$, $64$ or $128$.
For the transposed convolutional layers \cite{zeiler2010deconvolutional}, we set the Kernel Size to $2 \times 2$, Stride to 2, Padding to 1, Feature Map Number to $64$ or $32$.
The reconstruction network can be formulated as follows,


\begin{equation}
F^{rec}_i(z) = P(W_i * F^{rec}_{i-1}(z) + B_i), i\in\{1,2\}
\label{Eq::reci}
\end{equation}
where $z$ is the reconstruction signal input, and $P$ represents the sequential functions for activation and max-pooling.
We choose PReLU as the activation function in the Reconstruction Network.


\begin{equation}
            \begin{array}{rr}
             F^{rec}_5(z) = C(P(W_5 * F^{rec}_{4}(z) + B_5), F^{rec}_{2}(z)) &  \\
             F^{rec}_7(z) = C(P(W_7 * F^{rec}_{6}(z) + B_7), F^{rec}_{1}(z)) & 
            \end{array}
\label{Eq::recc}
\end{equation}
where $C$ denotes the concatenating function for jointing features.


After concatenating the features of the Residual Network and the Reconstruction Network, we calculate them with a convolutional layer of $1$ channel.
Then we obtain the final output $F_{out}(x, z)$ which is the same size as input.

%
%

\begin{table}[tbp]
\caption{Training parameters}
\label{tab::trainingParas}
\center
\begin{tabular}{l|l}
\hline
Parameters                     &   QP 37        \\
\hline
\multirow{1}{*}{Base Learning Rate}         &   $1e^{-4}$                 \\
\hline
\multirow{1}{*}{$\gamma$ Adjusting Coefficient}         &   $0.1$                \\
\hline
\multirow{1}{*}{Adjusting Epochs Interval}         &   $100$                \\
\hline
\multirow{1}{*}{Weight Decay}         &   $1e^{-4}$                \\
\hline
\multirow{1}{*}{Momentum}         &   $0.9$                \\
\hline
\multirow{1}{*}{Total Epochs}         &   $120$                \\
\hline
\end{tabular}
\end{table}

\subsection{Loss function, dataset and training}
\label{Subsec::training}

\textup{\textbf{Loss function}}.
We employ Mean Squared Error (MSE) \cite{wackerly2014mathematical} as the loss function for our proposed RRNet as follows,
\begin{equation}
L(\Theta) = \frac{1}{N}\sum^N_{i=1}||\Upsilon(Y_i|\Theta) - X_i||^2_2
\label{Eq::mse}
\end{equation}
where $\Theta$ encapsulates the whole parameter set of the network containing weights and bias and $\Upsilon(Y_i|\Theta)$ denotes the network module.
$X_i$ is a pixel of the original frame, where $i$ indexes each pixel.
$Y_i$ is the corresponding pixel of the reconstruction, that is compressed by HEVC when we turn off its deblocking and SAO.
$N$ is the number of pixels.

\textup{\textbf{Dataset}}.
We employ the DIV2K \cite{Ignatov_2018_ECCV_Workshops, agustsson2017ntire} dataset comprising $800$ training images and $100$ validating images of $2k$ resolution as the original frames.
Because modern video codecs operate on YUV color domain, we convert the original $900$ PNG images to YUV videos with FFMPEG \cite{FFmpeg} of GPU acceleration.
A modified HEVC reference software is then used to encode original frames to generate the reconstruction and residual with $QP22$, $QP27$, $QP32$, and $QP37$, respectively.
We finally extract $64\times64$ blocks from the Luma component of the reconstructed, residual, and original frames and use them as the inputs and labels for training our proposed RRNet.
In total, there are $522,939$ groups of inputs and labels for training and $66,650$ groups for validation.


\textup{\textbf{Training}}.
Once we obtain the residual and reconstruction patches of divided components, we feed them into the Residual Network and the Reconstruction Network, respectively, by batch-size of $16$.
Table~\ref{tab::trainingParas} exhibits the parameters of training procedure for $QP37$ samples.
We experiment with a larger learning ($1e^{-3}$) rate and a smaller learning rate ($1e^{-5}$), but the former one leads to the gradient explosion while the later one learns too slowly.
Therefore, $1e^{-4}$ is the appropriate base learning rate of $QP37$ model.
We adopt the Adaptive Moment Estimation (Adam) \cite{kingma2014adam} algorithm with the momentum of $0.9$ and the weight decay of $1e^{-4}$.
These parameter values are selected according to experience values.
When the model is trained less than $120$ epochs, the loss has not been convergent.
Accordingly, the $QP37$ model is trained with $120$ epochs.
After $100$ epochs, we decrease the learning rate by $10$ times.
After the $QP37$ model is derived, we fine tune it with $20$ epochs to obtain the other models: $QP22$, $QP27$, $QP32$.
Finally, we obtain the models for all the $QPs$ for testing.

\begin{table*}[tbp]
\caption{BD-rate of the SOTAs and proposed RRNet against HEVC under All Intra case}
\label{tab::comparisonRVHIntra}
\center
\begin{adjustbox}{max width=\textwidth}
\begin{tabular}{c|l|c|c|c|c}
\hline
\hline
{Class} & Sequence &   VRCNN \cite{dai2017convolutional} & EDSR Residual  & Partition-aware  & \ \ \ \ \ \ \ \ \  RRNet \ \ \ \ \ \ \ \ \ \\
 & &    vs. HEVC & Blocks \cite{lim2017enhanced} vs. HEVC & CNN \cite{lin2019partition} vs. HEVC & vs. HEVC\\
\hline
A & Traffic & $-8.1\% $& $-8.5\%$ & $-8.7\%$ & $\textbf{-10.2}\% $\\
 & PeopleOnStreet & $-7.7\% $ &$-7.8\%$& $-8.2\%$& $\textbf{-9.4}\% $\\
\hline
B & Kimono & $-5.9\% $ &$-6.6\%$& $-6.9\%$& $\textbf{-8.6}\% $\\
 & ParkScene & $-6.2\% $ &$-6.6\%$& $-6.9\%$& $\textbf{-8.1}\% $\\
 & Cactus & $-2.7\% $ &$-4.9\%$& $-5.4\%$& $\textbf{-5.8}\% $\\
 & BasketballDrive & $-5.2\% $ &$-4.6\%$& $-4.7\%$& $\textbf{-7.7}\% $\\
 & BQTerrace & $-2.9\% $ &$-2.9\%$& $-2.9\%$& $\textbf{-4.2}\% $\\
\hline
C & BasketballDrill & $-10.6\% $ &$-10.9\%$& $-11.3\%$& $\textbf{-13.8}\% $\\
 & BQMall & $-7.3\% $ &$-7.0\%$& $-7.4\%$& $\textbf{-9.3}\% $\\
 & PartyScene & $-4.6\% $ &$-4.5\%$& $-4.8\%$& $\textbf{-5.6}\% $\\
 & RaceHorses & $-5.8\% $ &$-5.0\%$& $-5.3\%$& $\textbf{-7.1}\% $\\
\hline
D & BasketballPass & $-7.6\% $ &$-7.3\%$& $-7.8\%$& $\textbf{-9.5}\% $\\
 & BQSquare & $-5.3\% $ &$-5.4\%$& $-5.8\%$& $\textbf{-6.3}\% $\\
 & BlowingBubbles & $-5.5\% $ &$-5.5\%$& $-5.7\%$& $\textbf{-6.7}\% $\\
 & RaceHorses & $-8.9\% $ &$-8.8\%$& $-9.1\%$& $\textbf{-10.2}\% $\\
\hline
E & FourPeople & $-10.0\% $ &$-10.4\%$& $-10.9\%$& $\textbf{-12.8}\% $\\
 & Johnny & $-9.1\% $ &$-8.1\%$& $-8.7\%$& $\textbf{-12.5}\% $\\
 & KristenAndSara & $-9.4\% $ &$-9.0\%$& $-9.6\%$& $\textbf{-11.8}\% $\\
\hline
& Class A & $-7.9\% $ &$-8.2\%$& $-8.5\%$& $\textbf{-9.8}\% $\\
 & Class B & $-4.6\% $ &$-5.1\%$& $-5.4\%$& $\textbf{-6.9}\% $\\
 & Class C & $-7.1\% $ &$-6.9\%$& $-7.2\%$& $\textbf{-8.9}\% $\\
 & Class D & $-6.8\% $ &$-6.7\%$& $-7.1\%$& $\textbf{-8.2}\% $\\
 & Class E & $-9.5\% $ &$-9.2\%$& $-9.7\%$& $\textbf{-12.4}\% $\\
\hline
Avg. & All & $-6.8\% $ &$-6.9\%$& $-7.2\%$& $\textbf{-8.9}\% $\\
\hline
\hline
\end{tabular}
\end{adjustbox}
\end{table*}

\section{Experimental results}
\label{Sec::experimental results}

To test the performance of the proposed algorithm, we embedded the proposed RRNet scheme into HEVC reference software as shown in Fig.~\ref{Fig::codingFrame}.
In this section, we first compare the proposed RRNet with VRCNN \cite{dai2017convolutional}, EDSR Residual Blocks \cite{lim2017enhanced}, Partition-aware CNN \cite{lin2019partition}, and HEVC on BD-rate \cite{Bjontegaard2001}, respectively.
Subsequently, we validate the multiple inputs function by comparing the dual-input residual and reconstruction with the solo input reconstruction.
Meanwhile, we compare the dual-input Residual and Reconstruction approach with the dual-input Partition and Reconstruction approach \cite{lin2019partition}.
Afterward, we evaluate the efficiency of different networks on the same inputs by comparing RRNet and EDSR Residual Blocks with the dual-input of residual and reconstruction.
For the test, we test all the sequences defined in HM-16.19 CTC \cite{KarlSharman2018} under the intra-coding and inter-coding configurations.

\begin{figure*}[tp]
\centering
\begin{minipage}[b]{.49\linewidth}
  \centering
  \includegraphics[width=\linewidth]{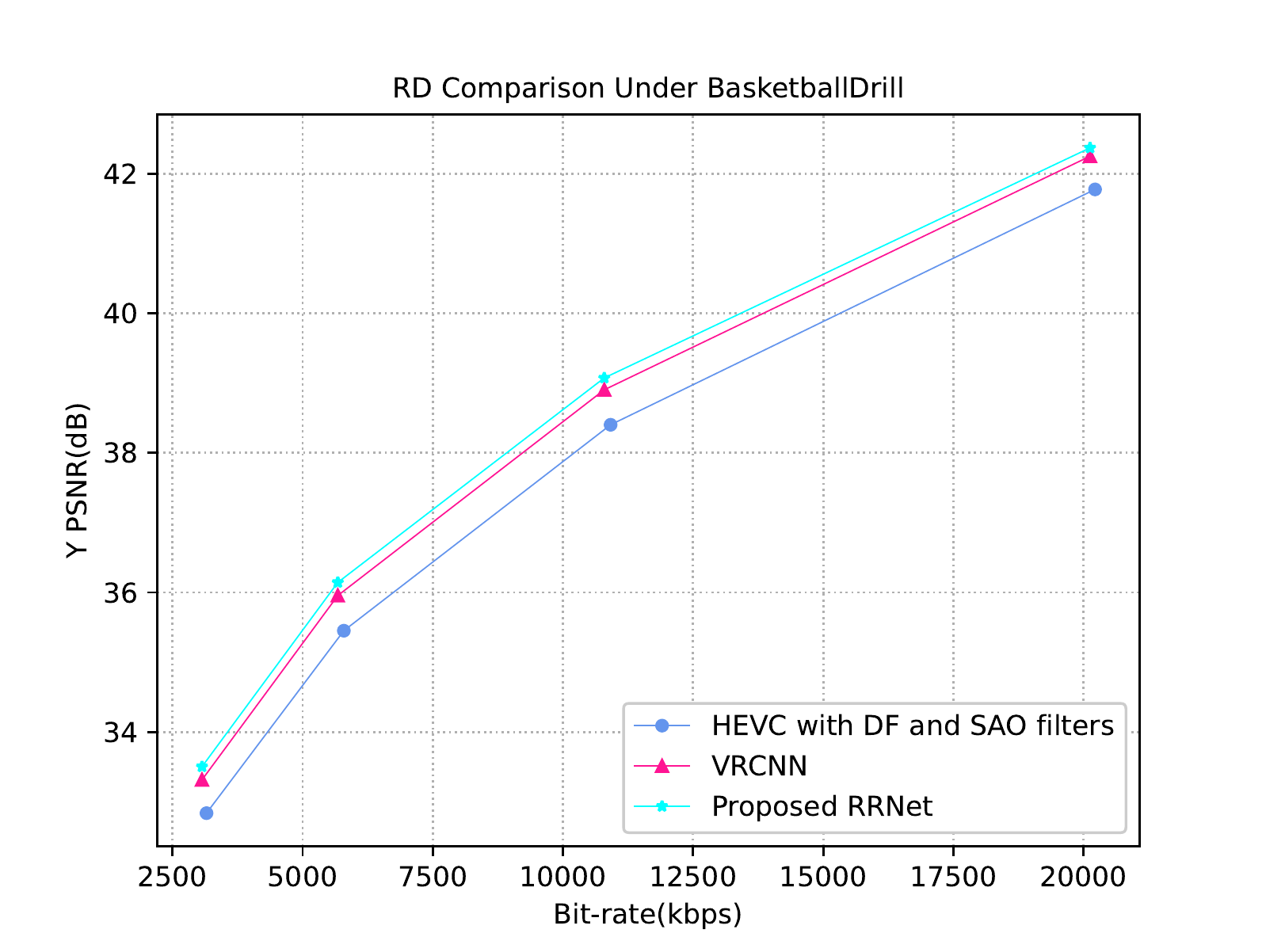}
  \centerline{(a) BasketballDrill }\medskip
\end{minipage}
\begin{minipage}[b]{.49\linewidth}
  \centering
  \includegraphics[width=\linewidth]{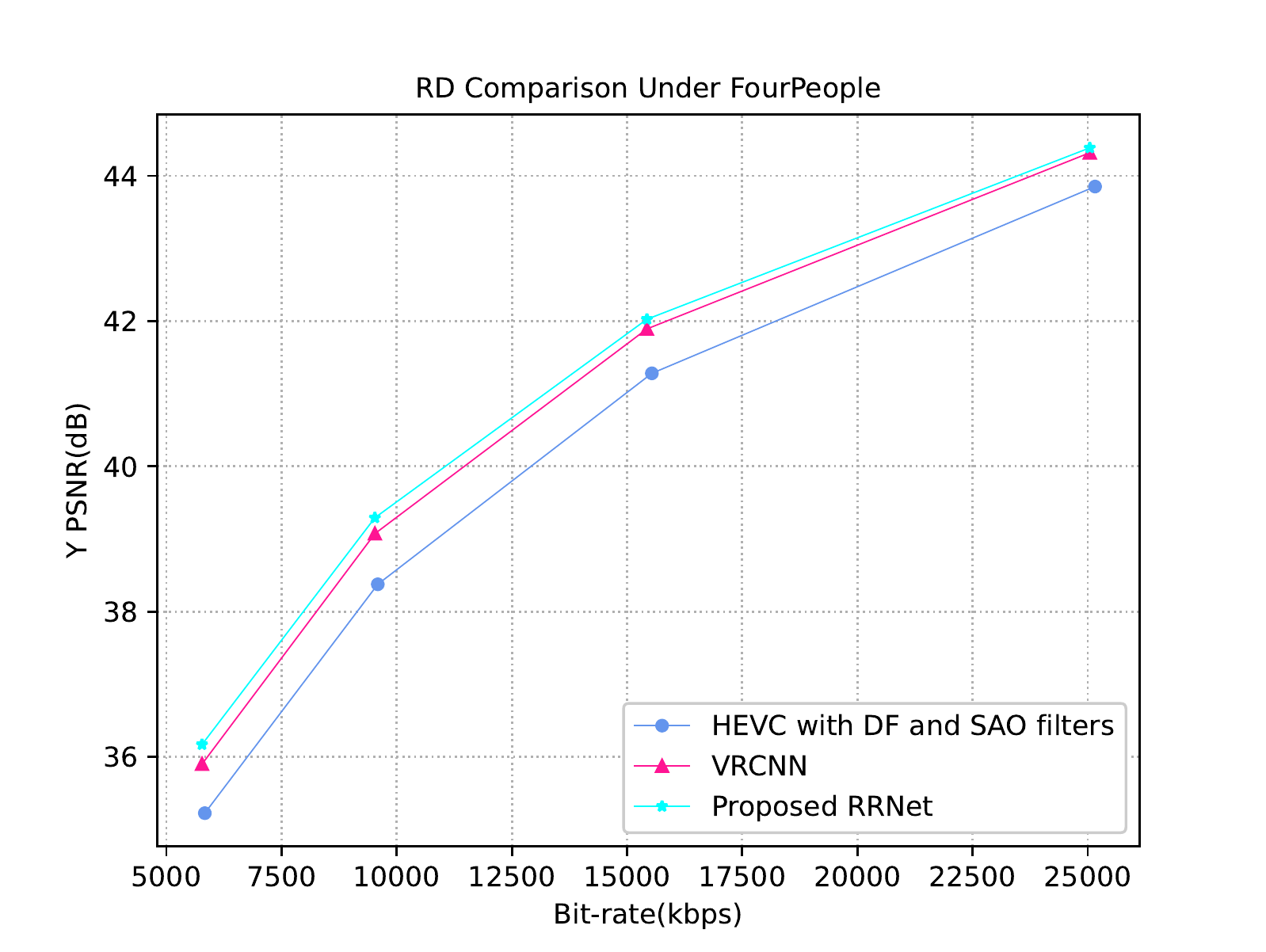}
  \centerline{(b) FourPeople }\medskip
\end{minipage}

\begin{minipage}[b]{.49\linewidth}
  \centering
  \includegraphics[width=\linewidth]{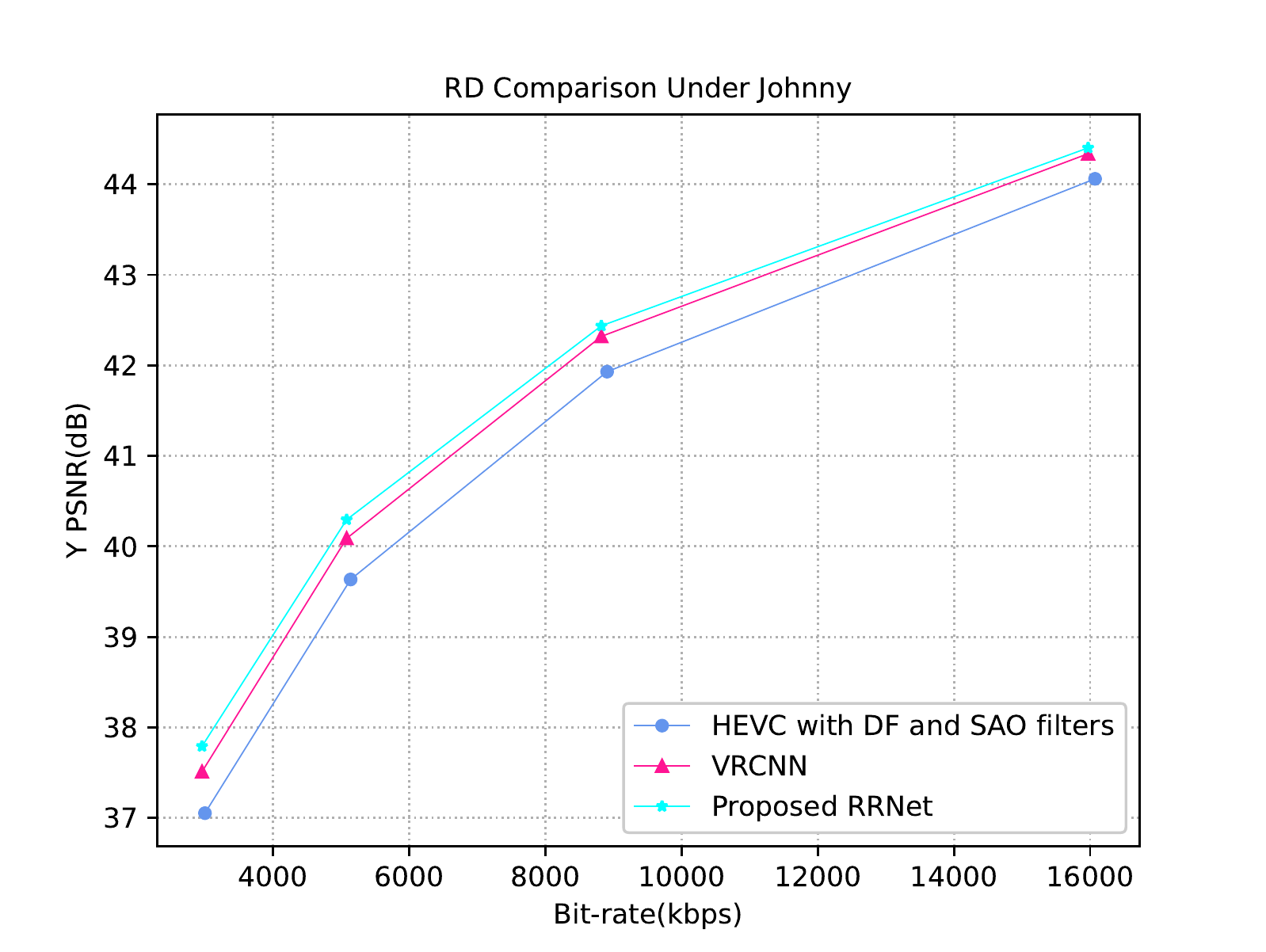}
  \centerline{(c) Johnny }\medskip
\end{minipage}
\begin{minipage}[b]{.49\linewidth}
  \centering
  \includegraphics[width=\linewidth]{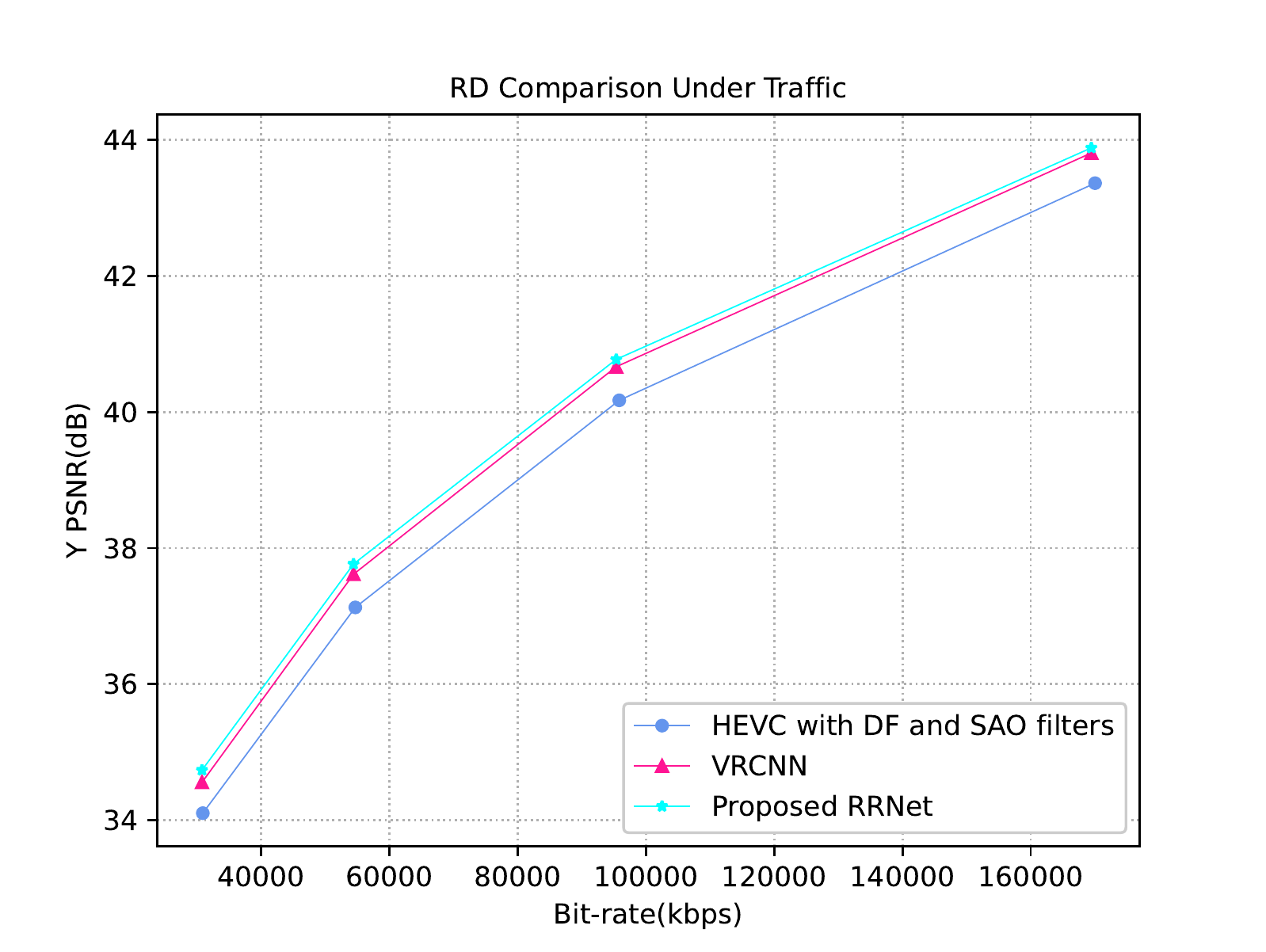}
  \centerline{(d) Traffic }\medskip
\end{minipage}
\caption{Comparison of RD curves in HEVC with DF and SAO, VRCNN and proposed RRNet on luminance.
The compared RD curves of BasketballDrill(a), FourPeople(b), Johnny(c) and Traffic(d) are shown.
It is obvious that our proposed RRNet outperforms HEVC with DF and SAO and VRCNN for all theses sequences under all tested QPs including $22, 27, 32 $ and $37$.
}
\label{fig:rd}
\end{figure*}

\begin{table}[tbp]
  \caption{The computational complexity of VRCNN and proposed RRNet against HEVC under All Intra case}
  \label{tab::complexity}
  \center
  \begin{tabular}{c|c|c|c}
  \hline
  \hline
  \multirow{1}{*}{Approches}         & Frame-work & Encoding Time & Decoding Time \\
  \hline
\multirow{1}{*}{VRCNN}         & Pytorch(C++) & $108.72\%$ & $420.41\%$  \\
  \hline
  \multirow{1}{*}{RRNet}         & Pytorch(C++) & $117.48\%$ & $1238.78\%$               \\
  \hline
  \hline
  \end{tabular}
\end{table}

\begin{table}[tp]
  \caption{BD-rate of VRCNN and proposed RRNet against HEVC under Random Access case}
  \label{tab::comparisonRVHInter}
  \center
  \begin{tabular}{c|l|c|c}
  \hline
  \hline
  {Class} & Sequence &  VRCNN vs. HEVC & RRNet vs. HEVC\\
  \hline
  A & Traffic & $-5.0\% $ & $\textbf{-6.0}\% $\\
   & PeopleOnStreet & $-1.4\% $ & $\textbf{-1.6}\% $\\
  \hline
  B & Kimono & $-1.9\% $ & $\textbf{-2.6}\% $\\
   & ParkScene & $-2.7\% $ & $\textbf{-3.4}\% $\\
   & Cactus & $-3.2\% $ & $\textbf{-3.9}\% $\\
   & BasketballDrive & $-1.4\% $ & $\textbf{-1.9}\% $\\
   & BQTerrace & $-5.2\% $ & $\textbf{-5.8}\% $\\
  \hline
  C & BasketballDrill & $-3.1\% $ & $\textbf{-4.3}\% $\\
   & BQMall & $-2.0\% $ & $\textbf{-2.5}\% $\\
   & PartyScene & $-0.5\% $ & $\textbf{-1.0}\% $\\
   & RaceHorses & $-1.3\% $ & $\textbf{-1.4}\% $\\
  \hline
  D & BasketballPass & $-0.7\% $ & $\textbf{-0.9}\% $\\
   & BQSquare & $-1.4\% $ & $\textbf{-2.1}\% $\\
   & BlowingBubbles & $-1.8\% $ & $\textbf{-2.4}\% $\\
   & RaceHorses & $-1.5\% $ & $\textbf{-1.6}\% $\\
  \hline
  E & FourPeople & $-8.2\% $ & $\textbf{-9.5}\% $\\
   & Johnny & $-7.6\% $ & $\textbf{-10.2}\% $\\
   & KristenAndSara & $-6.9\% $ & $\textbf{-7.6}\% $\\
  \hline
  & Class A & $-3.2\% $ & $\textbf{-3.8}\% $\\
   & Class B & $-2.9\% $ & $\textbf{-3.5}\% $\\
   & Class C & $-1.7\% $ & $\textbf{-2.3}\% $\\
   & Class D & $-1.4\% $ & $\textbf{-1.7}\% $\\
   & Class E & $-7.6\% $ & $\textbf{-9.1}\% $\\
  \hline
  Avg. & All & $-3.1\% $ & $\textbf{-3.8}\% $\\
  \hline
  \hline
  \end{tabular}
  \end{table}

\subsection{Performances of the proposed RRNet algorithm}
\label{subsec::performances}
Table~\ref{tab::comparisonRVHIntra} shows the comparison results of VRCNN \cite{dai2017convolutional}, EDSR Residual Blocks \cite{lim2017enhanced}, Partition-aware CNN \cite{lin2019partition}, and the proposed RRNet against HEVC under the all intra case.
Note that to ensure fairness, the EDSR Residual Blocks and Partition-aware CNN all employ eight convolutional layers, including three residual blocks as shown in Table~\ref{tab::gnet}, which have the same convolution layer depth as the one of the Residual Network in the proposed RRNet.
We train $QP37$ models of VRCNN, EDSR Residual Blocks, and Partition-aware CNN with $120$ epochs on the whole DIV2K dataset and then achieve the models of $QP32$, $QP27$ and $QP22$ by fine tuning the trained $QP37$ model with $20$ epochs.
These are identical to the process used to train RRNet as stated in Section~\ref{Subsec::training}.

We can see that the proposed RRNet algorithm outperforms VRCNN, EDSR Residual Blocks, and Partition-aware CNN by an average of $2.1\%$, $2.0\%$, and $1.7\%$, respectively.
Additionally, the RRNet method surpasses VRCNN, EDSR Residual Blocks, and Partition-aware CNN in every sequence in BD-rate.
Specifically, the proposed RRNet scheme outperforms VRCNN, EDSR Residual Blocks, and Partition-aware CNN by $2.9\%$, $3.2\%$, and $2.7\%$ on Class E, respectively.
Similarly, compared to the HEVC anchor, RRNet realizes a substantial gain on BD-rate with an average of $-8.9\%$.
The most remarkable individual difference occurs on BasketballDrill sequence with a gain of $-13.8\%$ on BD-rate.
This sequence contains particularly complex textures with very dramatic variations.
These performances demonstrate that RRNet effectively enhances the reconstruction by introducing the residual signal and developing customized networks for residual and reconstruction inputs.

Fig.~\ref{fig:rd} shows the luminance Rate-Distortion (RD) curves of the proposed RRNet approach, VRCNN, and HEVC anchor.
As illustrated, the PSNR of the proposed RRNet method is higher than the one of VRCNN and HEVC with in-loop filters under every QP in BasketballDrill, FourPeople, Johnny, and Traffic sequences.
This clearly shows that the proposed RRNet model is superior to the VRCNN and HEVC baseline approaches to enhance the quality of compressed video frames.

The time complexity \cite{Yiming2019} is exhibited in Table~\ref{tab::complexity}.
In all cases, we apply the same test environment.
Specifically, the GPU configuration is GTX 1080ti.
Due to the huge computation of CNN on the encoder side, VRCNN takes $8.72\%$ longer than HEVC.
Meanwhile, because of the dual-input networks, RRNet takes $17.48\%$ longer than HEVC.
On the decoder side, the results reflect a similar situation for complexity.
HEVC computes fastest while RRNet complexity overhead is $1238.78\%$.
We can adopt the methods of model compression and acceleration \cite{cheng2017survey,cheng2018model,cheng2018recent} to reduce the redundancy of the proposed RRNet model.
The solutions of model compression and acceleration includes parameter pruning, quantization, low-rank factorization, compact convolutional filters, and knowledge distillation.
We can use the parameter pruning and quantization based approaches to remove the redundancy of the RRNet parameters.
In addition, the low-rank factorization based methods are utilized to calculate the useful parameters of RRNet.
The compact convolutional filters are structurally designed to shrink the parameter space of RRNet and save computation and storage resources. 
The approaches based on knowledge distillation is used to train a more compact RRNet or learn a distilled RRNet model.

Table~\ref{tab::comparisonRVHInter} shows the experimental results in random access case. 
We can see that the proposed algorithm can bring an average of $-0.7\%$ and $-3.8\%$ BD-rate gain compared to VRCNN and HEVC, respectively. 
Again, we can also see that RRNet outperforms the other two methods in every class. 
Moreover, the peak difference between RRNet and VRCNN reaches $1.5\%$ on Class E. 
This demonstrates that the benefits brought by RRNet can be propagated to inter frames.
Thus the RRNet can bring significant performance improvements in random access case.

\begin{table}[tp]
\caption{The dual-input Residual and Reconstruction approach and the dual-input Partition and Reconstruction \cite{lin2019partition} approach versus Reconstruction only approach on the BD-rate}
\label{tab::comparisonRecoGNET}
\center
\begin{tabular}{l|c|c}
\hline
\hline
\multirow{1}{*}{}  & Partition and Reconstruction \cite{lin2019partition}& Residual and Reconstruction\\
 & vs. Reconstruction  & vs. Reconstruction\\
\hline
 Class A & $-0.4\% $ & $\textbf{-1.0}\% $\\
 Class B & $-0.2\% $ & $\textbf{-0.9}\% $\\
 Class C & $-0.4\% $ & $\textbf{-1.1}\% $\\
 Class D & $-0.4\% $ & $\textbf{-0.8}\% $\\
 Class E & $-0.6\% $ & $\textbf{-1.6}\% $\\
\hline
Avg. All & $-0.4\% $ & $\textbf{-1.0}\% $\\
\hline
\hline
\end{tabular}
\end{table}


\begin{table}[tp]
  \caption{The computational complexity of the dual-input Partition and Reconstruction method \cite{lin2019partition} and the dual-input Residual and Reconstruction approach against HEVC}
  \label{tab::complexityRRGG_PRHG}
  \center
  \begin{tabular}{c|c|c|c}
  \hline
  \hline
  \multirow{1}{*}{Approches}         & Frame-work & Encoding  & Decoding  \\
  & & Time & Time \\
  \hline
  \multirow{1}{*}{Partition Reconstruction \cite{lin2019partition}}         & Pytorch(C++) & $122.24\%$ & $1581.63\%$  \\
  \hline
  \multirow{1}{*}{Residual Reconstruction}         & Pytorch(C++) & $123.81\%$ & $1669.39\%$               \\
  \hline
  \hline
  \end{tabular}
\end{table}

\begin{table}[tbp]
\caption{Convolutional Parameters of EDSR Residual Blocks \cite{lim2017enhanced}}
\label{tab::gnet}
\center
\begin{tabular}{l|l}
\hline
\multirow{1}{*}{Kernel Size}         &   $3 \times 3$           \\
\hline
\multirow{1}{*}{Feature Map Number}         &   $32$                \\
\hline
\multirow{1}{*}{Stride}         &   $1$                \\
\hline
\multirow{1}{*}{Padding}         &   $1$                \\
\hline
\end{tabular}
\end{table}

\subsection{Results analysis of multiple inputs approaches}
\label{subsec::multipleInputs}
Here we compare the method with residual and reconstruction inputs to the method with only reconstruction input.
Additionally, we compare the dual-input Residual and Reconstruction approach with another multiple inputs approach that utilizes the mean mask of the PU partition \cite{lin2019partition} and Reconstruction.
Note to guarantee a fair comparison, all reconstruction sub-networks utilize the same network with eight convolutional layers, including three EDSR residual blocks shown in Table~\ref{tab::gnet}.

Table~\ref{tab::comparisonRecoGNET} exhibits the comparison of the dual-input Residual and Reconstruction scheme against Reconstruction only method and the comparison of the dual input PU Partition and Reconstruction method against Reconstruction only method.
On the one hand, the dual-input Residual and Reconstruction saves an average of $-1.0\%$ BD-rate compared with Reconstruction only method.
On the other hand, the dual-input Residual and Reconstruction method saves an average of $-0.6\%$ BD-rate over the dual input Partition and Reconstruction method. 
Specifically, the dual-input Residual and Reconstruction approach leads $-1.6\%$ BD-rate on Class E against the only Reconstruction method.
The peak difference of BD-rate between the dual-input Partition and Reconstruction method and the only Reconstruction method on Class E is $-0.6\%$.
In every class, the dual-input of the Residual and Reconstruction approach is better than the only Reconstruction method and the dual-input of the Partition and Reconstruction method on BD-rate.

These performances clearly show that based on the same network architecture for video reconstruction, the residual signal provides useful information for augmenting the quality.
This is reasonable because the inverse transformed residual provides the TU partition information and the detailed textures used to enhance the reconstruction.
Hence, introducing the residual signal augments the quality of the compressed video frame prominently.
In conclusion, compared to the only Reconstruction method and another multiple input methods based on the mean mask of the partition, the dual-input Residual and Reconstruction approach clearly augments the reconstruction. 
On the aspect of the time complexity, as shown in Table~\ref{tab::complexityRRGG_PRHG}, the dual-input Residual and Reconstruction approach and the dual-input Partition and Reconstruction method are approximately on the same level.

\begin{table}[tp]
\caption{BD-rate of RRNet against the dual-input Residual and Reconstruction with EDSR Residual Blocks \cite{lim2017enhanced}}
\label{tab::comparisonGGHA}
\center
\begin{tabular}{l|c}
\hline
\hline
\multirow{1}{*}{Class}         & RRNet vs. Residual and Reconstruction with EDSR Residual Blocks\\
\hline
 Class A & $-0.8\% $\\
 Class B & $-1.4\% $\\
 Class C & $-1.0\% $\\
 Class D & $-0.5\% $\\
 Class E & $-2.2\% $\\
\hline
Avg. All & $-1.2\% $\\
\hline
\hline
\end{tabular}
\end{table}

\begin{figure}[tp]
  \begin{center}
  \centering
  \includegraphics[width=.8\linewidth]{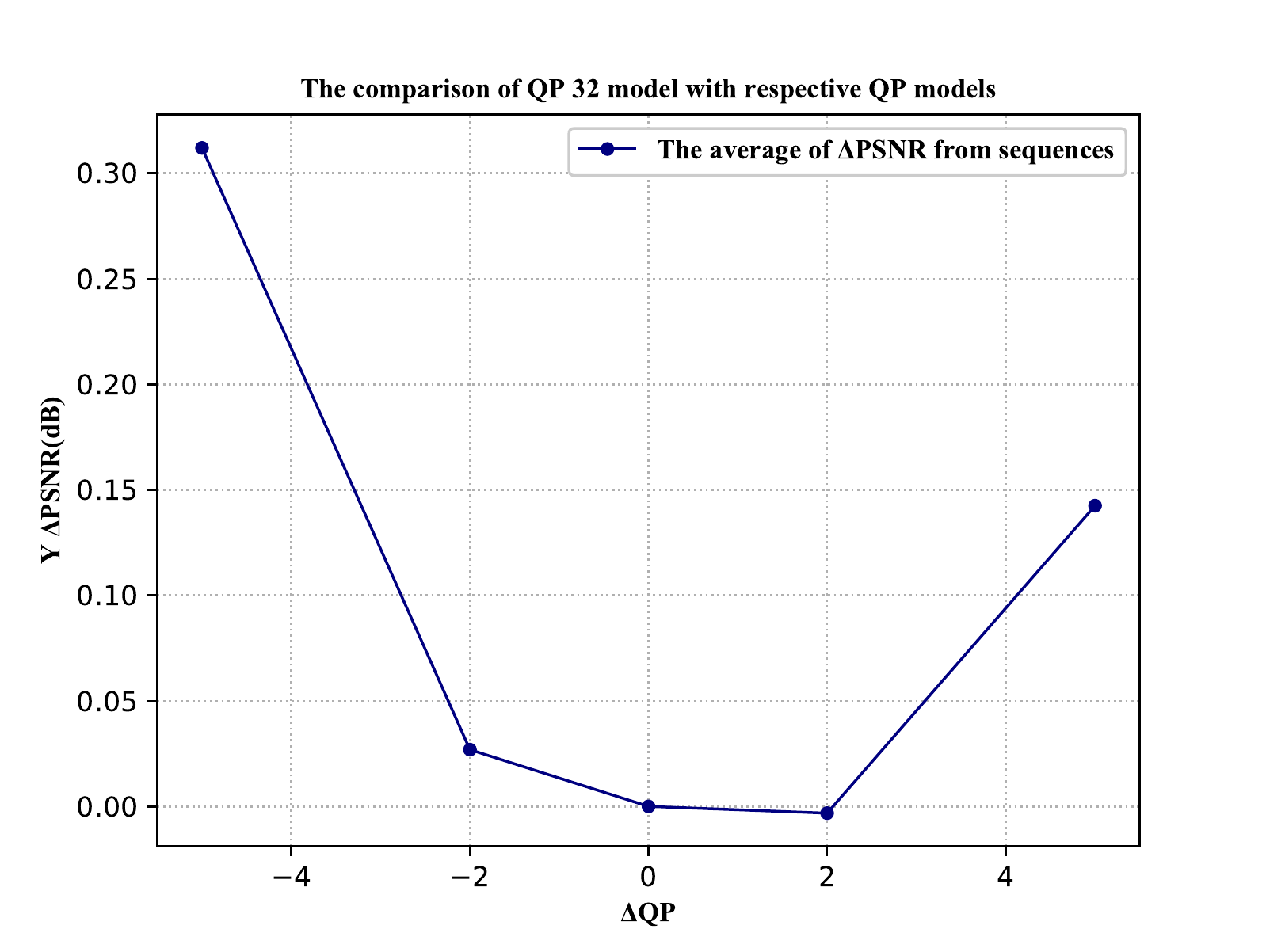}
  \caption{The comparison of QP $32$ model with respective QP models. The $\Delta$PSNR on $\Delta QP=0$ means the QP$32$ model compared to itself on PSNR is zero. Except QP$34$ setting, the PSNR of individual QP model is better than the one for the QP$32$ settings on the other QP parameters. The $\Delta$PSNR increases significantly with the absolute value of $\Delta$QP on the each side of $\Delta QP=0$.}
  \label{Fig::qpRange}
  \end{center}
\end{figure}

\begin{figure*}[tp]
  \centering
\includegraphics[angle=-90, scale=.16]{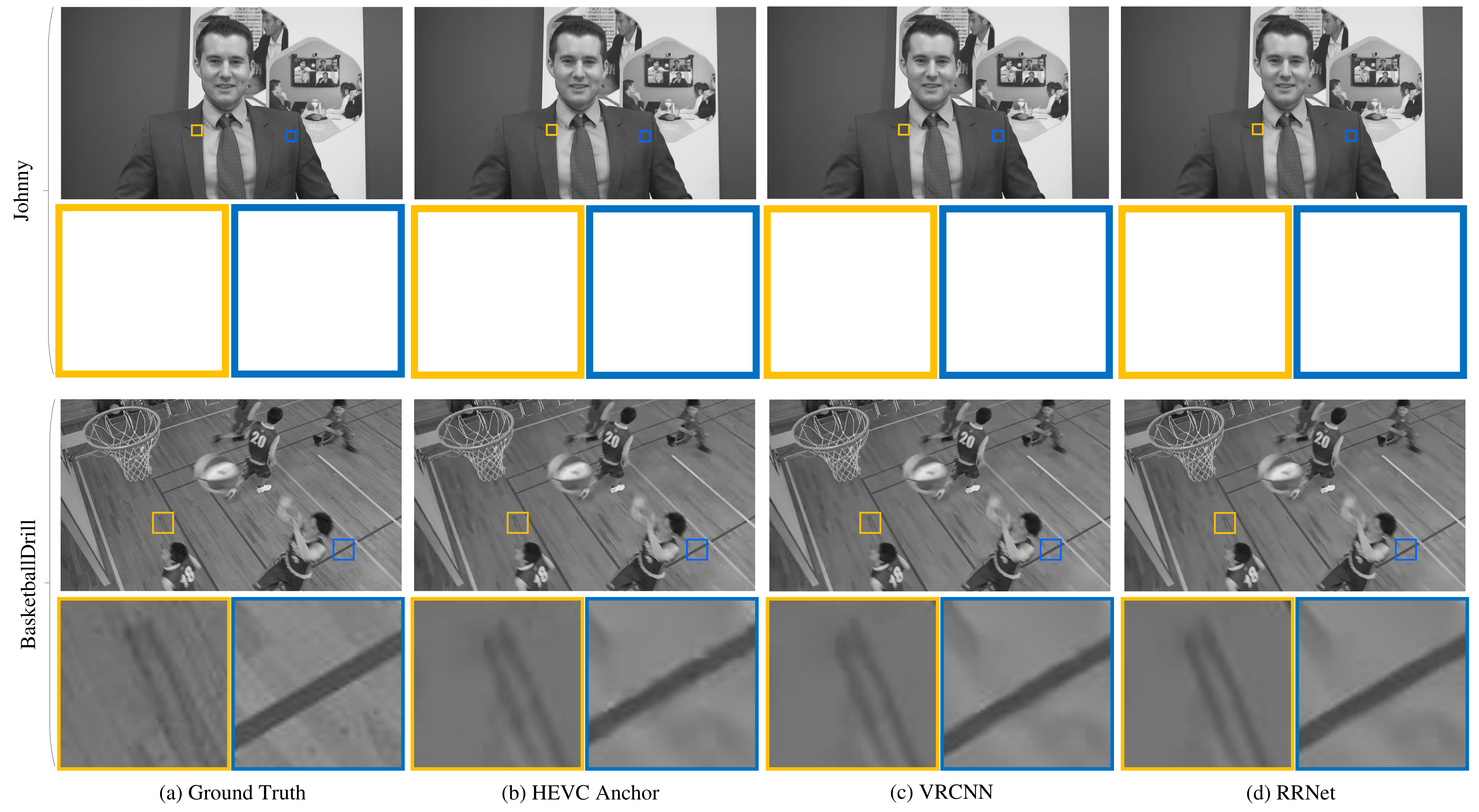}%
\caption{Visual comparisons between the ground truths, HEVC anchor, VRCNN, and proposed RRNet approach on the luminance of $QP37$ in Johnny and BasketballDrill sequences, respectively.
The groups of figures (a), (b), (c), and (d) are the original video, the video generated using HEVC, the video generated using VRCNN, the video generated using RRNet, respectively. 
(Zoom in for better visual effects.)
}
  \label{fig_sub}
\end{figure*}

\subsection{Results analysis of network architecture}
\label{subsec::architecture}
We compare the proposed RRNet approach with the dual-input of residual and reconstruction method with EDSR Residual Blocks to evaluate the performance of the proposed Residual Network and Reconstruction Network.
Note that both the RRNet and the second method have the same inputs.
The second method utilizes the EDSR Residual Blocks on both residual and reconstruction.
Table~\ref{tab::comparisonGGHA} shows the compared results between RRNet and the dual-input of residual and reconstruction approach with EDSR Residual Blocks.
RRNet gains an average of $-1.2\%$ BD-rate against the latter method.
Specifically, the proposed RRNet outperforms the dual-input of residual and reconstruction method with EDSR Residual Blocks in every class sequence for BD-rate.
The largest difference of BD-rate is $-2.2\%$ on the Class E sequence.
These demonstrate that both the Residual Network and the Reconstruction Network fit their respective signals very well.
The results also clearly demonstrate that processing the residual and reconstruction with unique architectures is beneficial.
Additionally, the validation of comparison provides evidence that the RRNet network shows an obvious improvement in the quality of coded frames.

\subsection{The performance from a specific QP model on different QPs}
\label{subsec::qpRange}
To validate the performance from an assigned QP model on other QP settings, as illustrated in Fig.~\ref{Fig::qpRange}, we compared the PSNR of QP$32$ when reconstructed by other QP models.
The $\Delta$PSNR on $\Delta QP=0$ means that the QP$32$ model compares itself on PSNR, and it should be zero.
Except for QP$34$, the PSNR of other QP models evaluated on itself is better than when it is evaluated on the QP$32$ model.
The $\Delta$PSNR increases dramatically with the absolute value of $\Delta$QP on both positive and negative sides.
Accordingly, specific QP tuned models outperform the other QP models when tuned for that specific setting. 
In summary, based on Fig.~\ref{Fig::qpRange}, a model can be reused to replace another model in the range of $-2$ to $2$ $\Delta$QP.



\subsection{Subjective Results}
\label{subsec::subjectiveRes}

Fig.~\ref{fig_sub} exhibits the visual comparisons between the ground truths, HEVC anchor, VRCNN, and proposed RRNet approach on the luminance of $QP37$ in Johnny and BasketballDrill sequences, respectively.
The groups of figures (a), (b), (c), and (d) are the original video, the video generated using HEVC, the video generated using VRCNN, the video generated using RRNet, respectively. 
In the Johnny, from the zoomed gold blocks, we can see that there are evident distortions and textures miss in the HEVC and VRCNN frames, while the RRNet frame shows smoother and more abundant textures.
We can see from the zoomed blue rectangles that the HEVC and VRCNN frames blur more severely than the RRNet frame.
From the BasketballDrill, we can see from the zoomed gold and blue blocks that the distortions in HEVC and VRCNN frames are more serious than the one of the RRNet frame.
The experimental results demonstrate that the proposed RRNet can bring better subjective qualities than the previous in-loop filtering methods.

\section{Conclusion}
\label{Sec::conclusion}
In this paper, we propose a new video deblocking solution that utilizing both reconstructed pixels as well as rich information and features available from the compression pipeline. The coding residual signal unique from compression pipeline is utilized as an additional input for improving the CNN based in-loop filter for HEVC.
In essence, it is introduced to enhance the quality of reconstructed compressed video frames. 
In this process, we first import the residual as an independent input to reinforce the textures and details.
Then, we custom designed RRNet approach that involves two separate CNNs: the Residual Network and the Reconstruction Network.
Each customized layer aims to reveal specific features that are characteristic of each type of frame.
In the Residual Network, we apply residual blocks to minimize the difference between the input frame and the output frame. 
In the Reconstruction Network, we utilize both downsampling and upsampling ladders to adapt to learn the features for the reconstruction frames.
The experimental results demonstrate that the proposed algorithms significantly reduce artifacts from both objective and subjective perspectives.
From the objective point of view, the BD-rate is significantly improved.
From the subjective point of view, the reconstruction quality of the compressed video frames is superior.
These results demonstrate that the proposed schemes improved the current state of the art significantly in BD rate reduction. In the future, we will try to create more advanced in-loop methods for video coding, while develop complexity reduction for the inference time model.

\ifCLASSOPTIONcaptionsoff
  \newpage
\fi

\bibliographystyle{IEEEtran}
\bibliography{predResi}

\end{document}